\documentclass[runningheads,a4paper]{llncs} 

\usepackage{makeidx}
\usepackage{pdfpages}
\usepackage{pgf,tikz}
\usepackage{mathrsfs}
\usetikzlibrary{arrows}
\usepackage{amssymb,amsmath,graphicx}
\usepackage{epstopdf}
\usepackage{multirow}
\usepackage{caption}

\begin{document}
\sloppy

\newcommand{\eps}{\varepsilon}
\newcommand{\real}{\mathbb{R}}
\newcommand{\nat}{{\mathbb{N}}}
\newcommand{\seq}[1]{\langle #1\rangle}

\newcommand{\crcl}{\mbox{$\mathcal{R}$}}
\newcommand{\bound}{\mbox{$\mathcal{L(R)}$} }
\newcommand{\ccc}{{\cal C}}
\newcommand{\VL}{\mbox{$\mathrm{VL}$}} 
\newcommand{\VR}{\mbox{$\mathrm{VR}$}}

%%%%%%%%%%%%%%%%%%%%%%%%%%%%%%%%%%%%%%%%%%%%%%%%%%%%%%%%%%%%%%%%%%%%%%%%%%%%%%%%%%%%%
\mainmatter
\title{A framework for small but rich vehicle routing problems.}

\titlerunning{A framework for the 2-VRP}

\author{Vladimir G.\ Deineko \inst{1} }

\institute{{Warwick Business School, Coventry, United Kingdom}}

\toctitle{A framework for small but rich}
\tocauthor{Vladimir Deineko}

\maketitle

%%%%%%%%%%%%%%%%%%%%%%%%%%%%%%%%%%%%%%%%%%%%%%%%%%%%%%%%%%%%%%%%%%%%%%%%%%%%%%%%%%%%%

%%%%%%%%%%%%%%%%%%%%%%%%%%%%%%%%%%%%%%%%%%%%%%%%%%%%%%%%%%%%%%%%%%%%%%%%%%%%%%%%%%%%%%%%%%%%%%

%%%%%%%%%%%%%%%%%%%%%%%%%%%%%%%%%%%%%%%%%%%%%%%%%%%%%%%%%%%%%%%%%%%%%%%%%%%%%%%%%%
%%%%%%%%%%%%%%%%%%%%%%%%%%%%%%%%%%%%%%%%%%%%%%%%%%%%%%%%%%%%%%%%%%%%%%%%%%%%%%%%%%
%\textsc{\title{{\bf On 2-period balanced TSP}} 
%\author{
%%Eranda \c{C}ela%%%
%%  \thanks{{\tt cela@opt.math.tu-graz.ac.at}.
%%  Institut f\"ur Optimierung und Diskrete Mathematik, TU Graz, Steyrergasse 30, A-8010 Graz, Austria}
%%  \and 
%  Vladimir Deineko%%%
%  \thanks{{\tt V.Deineko@warwick.ac.uk}.
%  Warwick Business School, Coventry, CV4 7AL, United Kingdom}
%  \thanks{Corresponding author; tel.\ +44 02476524501}
%%  \and Gerhard J.\ Woeginger%%%
%%  \thanks{{\tt gwoegi@win.tue.nl}.
%%  Department of Mathematics and Computer Science, TU Eindhoven, P.O.\ Box 513, 
%%  5600 MB Eindhoven, Netherlands} 
%}
%\date{}
%\maketitle
%}

%******************************************************************************
\begin{abstract}
In this paper we consider a 2-vehicle routing problem which can be viewed as a building block for the varieties of the vehicle routing problems (VRPs). To approach this problem, we suggest a framework based on the Held and Karp dynamic programming algorithm for the classical travelling salesman problem. An algorithm based on this framework shows an exceptionally good performance on published test data. Our approach can be easily extended to a variety of constraints/attributes in the VRP, hence the wording ``small but rich" in the title of our paper.

\medskip\noindent{\em Keywords.}
Combinatorial optimization; vehicle routing problem; dynamic programming; 2-period travelling salesman problem. 
\end{abstract}

%%%%%%%%%%%%%%%%%%%%%%%%%%%%%%%%%%%%%%%%%%%%%%%%%%%%%%%%%%%%%%%%%%%%%%%%%%%%%%%%%%
%%%%%%%%%%%%%%%%%%%%%%%%%%%%%%%%%%%%%%%%%%%%%%%%%%%%%%%%%%%%%%%%%%%%%%%%%%%%%%%%%%
\section{Introduction}
\label{sec1}
\nopagebreak

In the vehicle routing problem (VRP) a set of customers with  certain  requests are to be visited by vehicles. The vehicles are to be chosen from a fleet of heterogeneous vehicles, with various fixed and variable costs of usage. The objective is to find a minimum cost schedule for customer visits to deliver required services in the specified time and manner. The scheduling of visits may need to satisfy some additional constraints, e.g. with some visits demanded on particular days, or by particular vehicles, etc. 

It would not be an exaggeration to say that thousands of research papers devoted to the VRPs are published every year. As rightly mentioned by Michael Drexl \cite{Drexl}, this ``is certainly due to the intellectual challenge VRPs pose as well as to their relevance in logistics and transport''.

The VRP can be viewed as a combination of the two well known combinatorial optimization problems - the travelling salesman problem (see e.g. \cite{ABCC}) and the bin packing/knapsack problem (see e.g. \cite{KPP}). It is not surprising that the combination of these two problems creates new computational challenges for researchers and practitioners. For instance, the sizes of the TSP instances which are tractable by recently developed computational algorithms \cite{ABCC} are much bigger than the sizes of easy tractable VRP instances.

In many practical applications the number of customers visited by a single vehicle are not very large. We mention here just a few contextual examples: teachers visiting special needs pupils, nurses attending patients at home, food delivery in rural areas (long distances - hence few customers to visit), bulk deliveries of industrial goods (large items - hence few customers). The task of allocation of customers to vehicles adds a lot of complexity to the VRP, especially due to the various additional constraints arising in practice. The VRPs with multiple practice related constraints are referred to in recent publications as \emph {rich} VRPs (see surveys \cite{CAGR,Drexl,LKF}), or \emph{multi-attribute} VRPs \cite{VCGP1,VCGP2}.

In our study we decided to concentrate on the simplest possible version of the rich VRP - the VRP with many constraints/attributes but with only two vehicles, which we call the 2VRP. Further on, keeping in mind the practical applications mentioned above, we studied small (with not many customers) but rich VRPs (hence the title of the paper). We hope that better understanding of the 2VRP would allow researchers to develop new algorithms for the VRPs with many vehicles. A straightforward approach for the VRP with many vehicles could for instance be a heuristic which enumerates all possible pairs of vehicles and iteratively solves the 2VRPs.

The approach described in this paper is based on the well known Held and Karp dynamic programming approach \cite{HeldKarp}. 
%Unfortunately, exact solutions of the 2VRP can be obtained only for very small problems, say with no more than 20 customers.
We are not the first who suggest using  dynamic programming techniques for the VRPs (see e.g. \cite{DDS,GHKS,KMKS,Psaraft1,Psaraft2,VCGP2}). The way we use dynamic programming is different though to what has been suggested so far. We suggest a framework which permits to easily incorporate additional constraints and vary the objective function to optimise (see \cite{GHKS,Irnich,SMJM,SG,VCGP3} for frameworks suggested for various types of the VRPs). To address the \emph{curse of dimensionality} in dynamic programming, an aggregation scheme is included in the framework. 
We tested our framework on the 2-period balanced travelling salesman problem \cite{BassettoM11}. The results of computational experiments are impressive: for 60 benchmark instances suggested in \cite{BassettoM11}, better solutions have been  found for  57 instances.

\section{A framework for the 2VRP}
\subsection{Notations and definitions}

In this section we introduce a model for a vehicle routing problem (VRP) with only two vehicles (2VRP). We describe objects involved in the model and attributes associated with each object. In what follows we use an approach which finds optimal solutions for small size problems. To reduce the size of an initial problem we suggest making use of various aggregation techniques. As an example of a possible aggregation, one can think of considering all customers on a street as one customer with the demand defined as the sum of demands for the customers on that street.

There are two {\em not homogeneous} vehicles, vehicle 1 and vehicle 2, in the 2VRP. They have different capacities $W_1$ and $W_2$ and different costs of travelling. The vehicles will be used for delivering demanded goods to customers.

We assume that each customer is located in an estate with only two entry points. A network of one-way roads within the estate connects these two points. So the travel costs within the estate are asymmetric costs.
To distinguish between two entry points to the estate, we refer to  one of the points as the {\em left node} and denote it by $L(i)$. The other of the two points is  referred to as the {\em right node}, and is denoted by $R(i)$. 

We assume that customer $i$ in our problem has a set of seven attributes
$\{L(i),\ R(i),\ l^1_L(i),\ l^1_R(i),\ l^2_L(i),\ l^2_R(i),\ w(i)\}$ with the following meaning:
\begin{itemize}
\item the left node $L(i)$;
\item the right node $R(i)$;
\item cost of travelling from the left node to the right node $l^m_L(i)$, if travelled by vehicle $m$, $m=1,2$;
\item cost of travelling from the right node to the left node $l^m_R(i)$, if travelled by vehicle $m$, $m=1,2$;
\item the demand $w(i)$.  
\end{itemize}

So, given a set of $n$ customers $\{1,2,\ldots,n\}$, with the demands $w(i)$, they have to be visited by one of the two vehicles which delivers the demanded goods from depots to the customers. The total demand of all customers is less than the total capacity of the two vehicles, so only one route per vehicle is needed. The vehicles may travel from different depots and return to different depots.  Assume that vehicle 1 travels from depot $d^1_R$ and returns to depot $d^1_L$, and vehicle 2 travels from depot $d^2_R$ and returns to depot $d^2_L$.

The costs of travelling between the nodes are given by the two $2(n+4)\times 2(n+4)$ cost matrices $C^1=(c^1_{ij})$ and $C^2=(c^2_{ij})$,  where $c^m(i,j)$ is the cost of travelling by vehicle $m$ from node $i$ to node $j$. 

%These matrices represent the travel costs between nodes, however, 
A visit of customer $i$ adds to the travel costs either cost $l^m_L(i)$, or cost $l^m_R(i)$ depending on the way the customer is visited - from the left entry node, or from the right entry node.

Our approach will utilise the well known dynamic programming algorithm for the travelling salesman problem. So we would like to view routes for the two vehicles as one route. To simplify further considerations, we introduce an auxiliary customer $0$ with the set of attributes $\{d^1_L,d^2_R,0,\infty,0,\infty,0\}$. The reason for introduction of the auxiliary customer $0$ and placing it into the two-vehicle route is to separate points visited by the two different vehicles. Cost of travelling from $d^1_L$ to $d^2_R$ costs nothing, while travelling in opposite direction is forbidden (by the infinitely large cost).

The two-vehicle route starts from the node $d^1_R$, visits all customers  from the set $U=\{0\}\cup \{1,2,\ldots,n\}$, and ends in the node $d^2_L$. Vehicle 1 is the first to start visiting customers in the route. Visiting customer $0$ means that the mode of travelling is changed from travelling by vehicle 1 to travelling by vehicle 2. Vehicle 2 will travel from customer $0$, i.e.\ from depots $d^2_R$, to depot $d^2_L$.

The objective is to find a minimum cost route  of delivering the requested demand to all customers. The total demand of customers visited by each vehicle should not exceed the corresponding vehicle's capacity.

\subsection{Dynamic programming recursions}
In this section, the well known Held \& Karp \cite{HeldKarp} dynamic programming algorithm for the TSP is adapted for the case of the 2VRP formulated above.

Let $J$ be a subset of customers not containing $i$, so $J\subset U$, $i \notin J$.
Denote as  $\VL[i,J]$ the minimum cost of an optimal 2-vehicle route among all routes which start from a visit of customer $i$ from the corresponding left node, then visit all the customers in set $J$, and stop in depot $d^2_R$. Similarly, define $\VR[i,J]$ to be the cost of the optimal route that starts visiting customer $i$ from the corresponding right node, and visits then the customers in $J$. The optimal cost of a 2-vehicle tour can be calculated as
 
\begin{equation}
\label{eq:dp0}
V=min_{i\in U\setminus \{0\}}\{c^1_{d^1_R,L(i)}+\VL[i,U\setminus \{i\}],c^1_{d^1_R,R(i)}+\VR[i,U\setminus \{i\}]\}.
\end{equation}

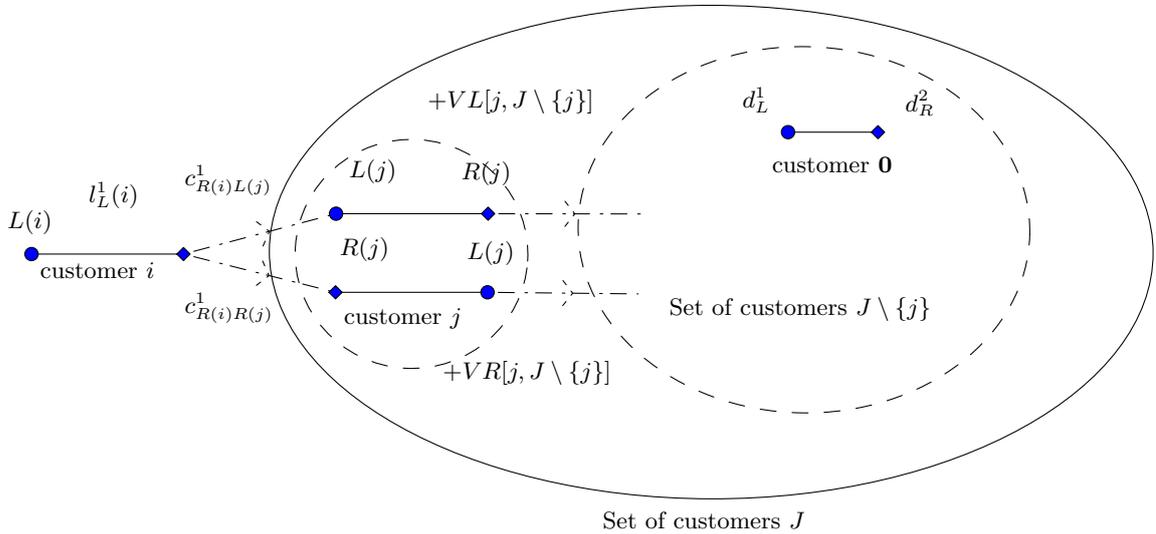
\begin{figure}
\unitlength=1cm
%\begin{center}
\definecolor{qqqqff}{rgb}{0.,0.,1.}
\begin{tikzpicture}[line cap=round,line join=round,>=triangle 45,x=1.0cm,y=1.0cm]
\clip(1.55,1.11) rectangle (17.,10.3);
\draw (2.,6.)-- (4.,6.);
\draw (6.007042253521121,6.535211267605635)-- (8.00704225352112,6.535211267605635);
\draw (6.,5.492957746478873)-- (8.,5.492957746478873);
\draw [rotate around={-3.576334374995288:(7.,6.001760563380284)},dash pattern=on 6pt off 6pt] (7.,6.001760563380284) ellipse (1.5307841759435992cm and 1.5179812710839782cm);
\draw [rotate around={-1.2364076028561437:(12.163732394366187,6.32218309859155)},dash pattern=on 6pt off 6pt] (12.163732394366187,6.32218309859155) ellipse (2.971769702905846cm and 2.428082112076715cm);
\draw [rotate around={-0.1469119331988365:(10.943661971830982,6.026408450704226)}] (10.943661971830982,6.026408450704226) ellipse (5.81427438136858cm and 3.271811342548349cm);
\draw (9.4,2.7) node[anchor=north west] {Set of customers $J$};
\draw (10.278169014084503,5.545774647887325) node[anchor=north west] {Set of customers $J \setminus \{j\}$};
\draw (6.,5.4) node[anchor=north west] {customer $j$};
\draw (11.95422535211267,7.616197183098592)-- (13.137323943661965,7.616197183098592);
\draw (11.239436619718305,8.330985915492958) node[anchor=north west] {$d^1_L$};
\draw (13.383802816901403,8.28169014084507) node[anchor=north west] {$d^2_R$};
\draw (11.63,7.4) node[anchor=north west] {customer $\mathbf{0}$};
\draw (3.9,7.3) node[anchor=north west] {$c^1_{R(i)L(j)}$};
\draw (3.9,5.6) node[anchor=north west] {$c^1_{R(i)R(j)}$};
\draw [dash pattern=on 1pt off 3pt on 6pt off 4pt] (4.,6.)-- (6.007042253521121,6.535211267605635);
\draw [dash pattern=on 1pt off 3pt on 6pt off 4pt] (5.128553288839242,6.300947543690467) -- (5.046389296616109,6.106849996844511);
\draw [dash pattern=on 1pt off 3pt on 6pt off 4pt] (5.128553288839242,6.300947543690467) -- (4.960652956905013,6.428361270761123);
\draw [dash pattern=on 1pt off 3pt on 6pt off 4pt] (4.,6.)-- (6.,5.492957746478873);
\draw [dash pattern=on 1pt off 3pt on 6pt off 4pt] (5.125433196241976,5.714678907994991) -- (4.959114330399999,5.585207620928324);
\draw [dash pattern=on 1pt off 3pt on 6pt off 4pt] (5.125433196241976,5.714678907994991) -- (5.0408856696,5.907750125550549);
\draw [dash pattern=on 1pt off 3pt on 6pt off 4pt] (8.00704225352112,6.535211267605635)-- (10.031690140845065,6.531690140845072);
\draw [dash pattern=on 1pt off 3pt on 6pt off 4pt] (9.14876740994213,6.533225658637946) -- (9.019076852856427,6.367077716392306);
\draw [dash pattern=on 1pt off 3pt on 6pt off 4pt] (9.14876740994213,6.533225658637946) -- (9.01965554150976,6.699823692058401);
\draw [dash pattern=on 1pt off 3pt on 6pt off 4pt] (8.,5.492957746478873)-- (9.982394366197179,5.471830985915494);
\draw [dash pattern=on 1pt off 3pt on 6pt off 4pt] (9.120591243742526,5.481015388641475) -- (8.989424211955537,5.316030573940693);
\draw [dash pattern=on 1pt off 3pt on 6pt off 4pt] (9.120591243742526,5.481015388641475) -- (8.992970154241641,5.648758158453672);
\draw (7.1,8.3) node[anchor=north west] {$+VL[j,J\setminus\{j\}]$};
\draw (7.3,4.7) node[anchor=north west] {$+VR[j,J\setminus\{j\}]$};
\draw (6.063380281690138,7.369718309859155) node[anchor=north west] {$L(j)$};
\draw (7.542253521126757,7.345070422535212) node[anchor=north west] {$R(j)$};
\draw (5.94014084507042,6.334507042253522) node[anchor=north west] {$R(j)$};
\draw (7.616197183098588,6.285211267605635) node[anchor=north west] {$L(j)$};
\draw (2.,6.) node[anchor=north west] {customer $i$};
\draw (1.57,6.7) node[anchor=north west] {$L(i)$};
\draw (2.637323943661972,7.073943661971831) node[anchor=north west] {$l^1_L(i)$};
\begin{scriptsize}
\draw [fill=qqqqff] (2.,6.) circle (2.5pt);
\draw [fill=qqqqff] (4.,6.) ++(-2.5pt,0 pt) -- ++(2.5pt,2.5pt)--++(2.5pt,-2.5pt)--++(-2.5pt,-2.5pt)--++(-2.5pt,2.5pt);
\draw [fill=qqqqff] (6.007042253521121,6.535211267605635) circle (2.5pt);
\draw [fill=qqqqff] (8.00704225352112,6.535211267605635) ++(-2.5pt,0 pt) -- ++(2.5pt,2.5pt)--++(2.5pt,-2.5pt)--++(-2.5pt,-2.5pt)--++(-2.5pt,2.5pt);
\draw [fill=qqqqff] (6.,5.492957746478873) ++(-2.5pt,0 pt) -- ++(2.5pt,2.5pt)--++(2.5pt,-2.5pt)--++(-2.5pt,-2.5pt)--++(-2.5pt,2.5pt);
\draw [fill=qqqqff] (8.,5.492957746478873) circle (2.5pt);
\draw [fill=qqqqff] (11.95422535211267,7.616197183098592) circle (2.5pt);
\draw [fill=qqqqff] (13.137323943661965,7.616197183098592) ++(-2.5pt,0 pt) -- ++(2.5pt,2.5pt)--++(2.5pt,-2.5pt)--++(-2.5pt,-2.5pt)--++(-2.5pt,2.5pt);
\end{scriptsize}
\end{tikzpicture}
\vspace{-0.4cm}
\caption{Illustration of the calculation of length $\VL[i,J]$, the shortest route from customer $i$ through all customers in set $J$ (the case of $\mathbf{0}\in J$.)}
\end{figure}

We assume here that the total demand from customers is bigger than the capacity of each of the vehicles, so in the formula above $0$ cannot be the first customer. 
Values $\VL[i,J]$ and $\VR[i,J]$ for all customers $i$ and subsets $J$, $J\subset \{0\}\cup\{1,2,\ldots,n\}$, are calculated as shown in the recursions below:

\begin{equation}
\label{eq:dp1}
\begin{aligned}%{2}
\VL[i,J]& =%\left\{
\begin{cases}
\left.
\min_{j\in J}
\begin{cases}
				l^1_L(i)+c^1_{R(i),L(j)}+\VL[j,J\setminus\{j\}] \\
				l^1_L(i)+c^1_{R(i),R(j)}+\VR[j,J\setminus\{j\}]
\end{cases}
\hspace{-2.0ex}\right\}
&\mbox{if }0\in J, \\
\\
\left.
\min_{j\in J}
\begin{cases}
				l^2_L(i)+c^2_{R(i),L(j)}+\VL[j,J\setminus\{j\}]\\
				l^2_L(i)+c^2_{R(i),R(j)}+\VR[j,J\setminus\{j\}]
\end{cases}
\hspace{-2.0ex}\right\}
&\mbox{if }0\notin J, w(\{i\}\cup J)\le W_2\\
\infty & \mbox{otherwise}
\end{cases}
%\right.
\end{aligned}
\end{equation}

\begin{equation}
\begin{aligned}%{2}
\VR[i,J]& =%\left\{
\begin{cases}
\left.
\min_{j\in J}
\begin{cases}
				l^1_R(i)+c^1_{L(i),L(j)}+\VL[j,J\setminus\{j\}] \\
				l^1_R(i)+c^1_{L(i),R(j)}+\VR[j,J\setminus\{j\}]
\end{cases}
\hspace{-2.0ex}\right\}
&\mbox{if }0\in J, \\\\%[-1.0ex]
\left.
\min_{j\in J}
\begin{cases}
				l^2_R(i)+c^2_{L(i),L(j)}+\VL[j,J\setminus\{j\}]\\
				l^2_R(i)+c^2_{L(i),R(j)}+\VR[j,J\setminus\{j\}]
\end{cases}
\hspace{-2.0ex}\right\}
&\mbox{if }0\notin J, w(\{i\}\cup J)\le W_2\\
\infty &\mbox{otherwise}
\end{cases}
%\right.
\end{aligned}
\end{equation}

\begin{equation}
\begin{aligned}%{2}
\VR[0,J]& =%\left\{
\begin{cases}
\left.
\min_{j\in J}
\begin{cases}
				c^2_{R(0),L(j)}+\VL[j,J\setminus\{j\}\\
				c^2_{R(0),R(j)}+\VR[j,J\setminus\{j\}
\end{cases}
\hspace{-2.0ex}\right\}
&\mbox{if }
\begin{cases}w(U\setminus (\{i\}\cup J))\le W_1,\\ 
\mbox{and }w(\{i\}\cup J)\le W_2
\end{cases}\\
\infty & \mbox{otherwise}
\end{cases}
%\right.
\end{aligned}
\end{equation}

The boundary conditions are:
\begin{equation}
\label{eq:dpB}
\begin{aligned}%{2}
\VL[i,\emptyset]& =%\left\{
				l^2_R(i)+c^2_{L(i),L(0)},
\\
\VR[i,\emptyset]& =%\left\{
				l^2_L(i)+c^2_{R(i),L(0)}.
\end{aligned}
\end{equation}

Recursions (\ref{eq:dp0})-(\ref{eq:dpB}) extend Held \& Karp recursions to the case of 2VRP: since there are only two vehicles, the capacity constraints are easily verified without any extra dimensions or complicated calculations. Notice that we use notation $w(J)$ for the sum of demands of all items in set $J$. For instances with relatively small sizes, these days computers perform calculations (\ref{eq:dp0})-(\ref{eq:dpB}) within seconds. However when the number of customers approaches $20$, the computations on a standard laptop become more and more problematic. To make the dynamic programming approach practical, we suggest an aggregation strategy described in the next section.

\subsection{Aggregation strategies and local search}

\begin{figure}
\unitlength=1cm
\begin{center}
\begin{picture}(11.5,7.5)
%\put(-0.5,0)
{
\begin{picture}(10.5,6.5)
\includegraphics[scale=0.8]{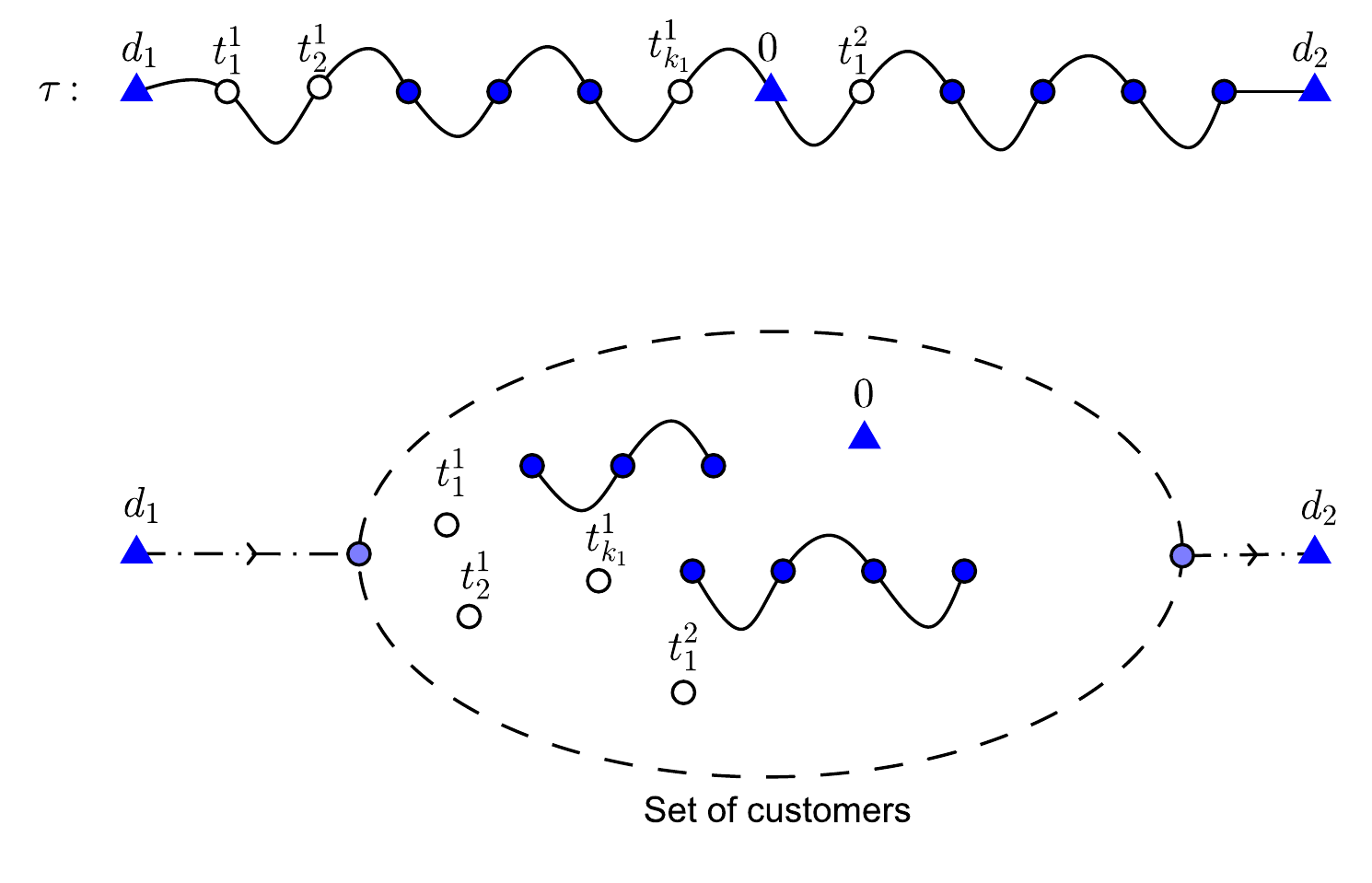}% \hspace{0.5cm}
\end{picture} 
}
\end{picture}
%\put(8.8,0){\makebox(0,0)[cc]{$B$}}
\end{center}
\caption{ Illustration of the ``sliding subsets" heuristic: first step of disassembling; each subset contains $2$ customers.}
\label{fig:disassembling1}
\end{figure}

As was mentioned above, the dynamic programming approach can be used for small size problems but may become impractical in real life applications. In this section we describe an approach for data aggregation  and   reducing the size of the initial problem. 

We start with a feasible 2VRP solution with $n$ customers. We ``cut" this solution into a small number of sub-paths. One can think of various possible ways of cutting a 2-VRP solution into sub-paths.
Each subpath is replaced then by a new customer. The way we describe the customers in the 2VRP in the previous section permits an easy replacement of any subpath by a new customer. A left node for the new customer is the first node  in the subpath (i.e. either left or right node of the first customer); the right node of the new customer is the last node in the subpath; the demand of the new customer is the sum of the demands of customers in the sub-path; the left/right lengths are calculated as the corresponding lengths of the subpath.

Obviously, an exact solution obtained for a new small-size problem can be viewed only as an approximate solution for the initial 2-VRP. This approximate solution can be again ``disassembled" into a small number of sub-paths, and the process of solving small size problems is repeated again, until  all possible ways of disassembling and aggregating a solution have been enumerated and no further improvement was achieved.

We suggest the following straightforward approach, which we call the sliding sub-sets method. Assume that we have an initial solution to the 2VRP: $\tau=\seq{d_1,t^1_1,t^1_2,\ldots,t^1_{k_1},0,t^2_1,t^2_2,\ldots,t^2_{k_2},d_2}$. Here vehicle one's route is 
$\tau=\seq{d_1,t^1_1,t^1_2,\ldots,t^1_{k_1},0}$ and vehicle two's route is 
$\seq{0,t^2_1,t^2_2,\ldots,t^2_{k_2},d_2}$.

We disassemble this solution into a new set of customers for the new 2VRP according to the following procedure. First, define $0$ as a customer in the new 2VRP and delete it from $\tau$.
Let $s$ now be a small constant, a parameter of the algorithm. Choose two subsets of customers containing $s$ items each so that the customers in each subset are picked from consecutive positions in $\tau$. Subset $S_1$ will always contain at least one customer from vehicle $1$'s route, and $S_2$ contains at least one customer from vehicle $2$'s route. On the first disassembling step  define the first subset as $S_1=\{t^1_1,\ldots,t^1_s\}$, and the second subset as $S_2=\{t^1_{k_1-s+2},\ldots,t^1_{k_1},t^2_1\}$. Notice that subset $S_2$ is chosen to ensure that at least one customer from vehicle 2's route is included in the subset. Delete $S_1$ and $S_2$ from $\tau$ and add them to the set of customers in the new 2VRP. Consider $d_2$ as a new customer, delete it from $\tau$. Sub-paths which are left in $\tau$ are considered as  aggregated customers and added to the new 2VRP. On the first step these are: depot $d_1$, sub-path $\seq{t^1_{s+1},\ldots,t^1_{k_1-s+1}}$, and the sub-path
$\seq{t^2_2,\ldots,t^2_{k_2}}$. So the new 2VRP contains $2s +5$ customers (depots are counted here as customers). 

Figure \ref{fig:disassembling1} illustrates the concept of sliding subsets. 
To simplify drawings the customers are depicted as points, however we treat them rather as segments, as described in the previous section. On the other hand, the nodes in the tour are not connected by straight lines, as one would expect. Recall that in our model we view customers as ``paths", therefore curves are chosen to illustrate the connections. 

If no better solution is found, we redefine subset $S_2$ by deleting, say, $l$ first elements and adding $l$ new elements ($l$ is a parameter, to which we refer as the {\em step}). So subset $S_2$ slides along the tour. We repeat the process until we reach the end of the tour. Then we change (slide) set $S_1$ (with the step $l$) and redefine set $S_2$ to follow set $S_1$ similar to the first step settings described above.

If the solution of the 2VRP was improved, the process of disassembling is applied to the new solution. The process is stopped when all possible positions for subsets $S_1$ and $S_2$ are considered and no improvements found.

\begin{figure}
\unitlength=1cm
\begin{center}
%\begin{picture}(11.5,8.5)
%\put(-0.5,0)
{
\begin{picture}(10.5,8.6)
\includegraphics[scale=0.8]{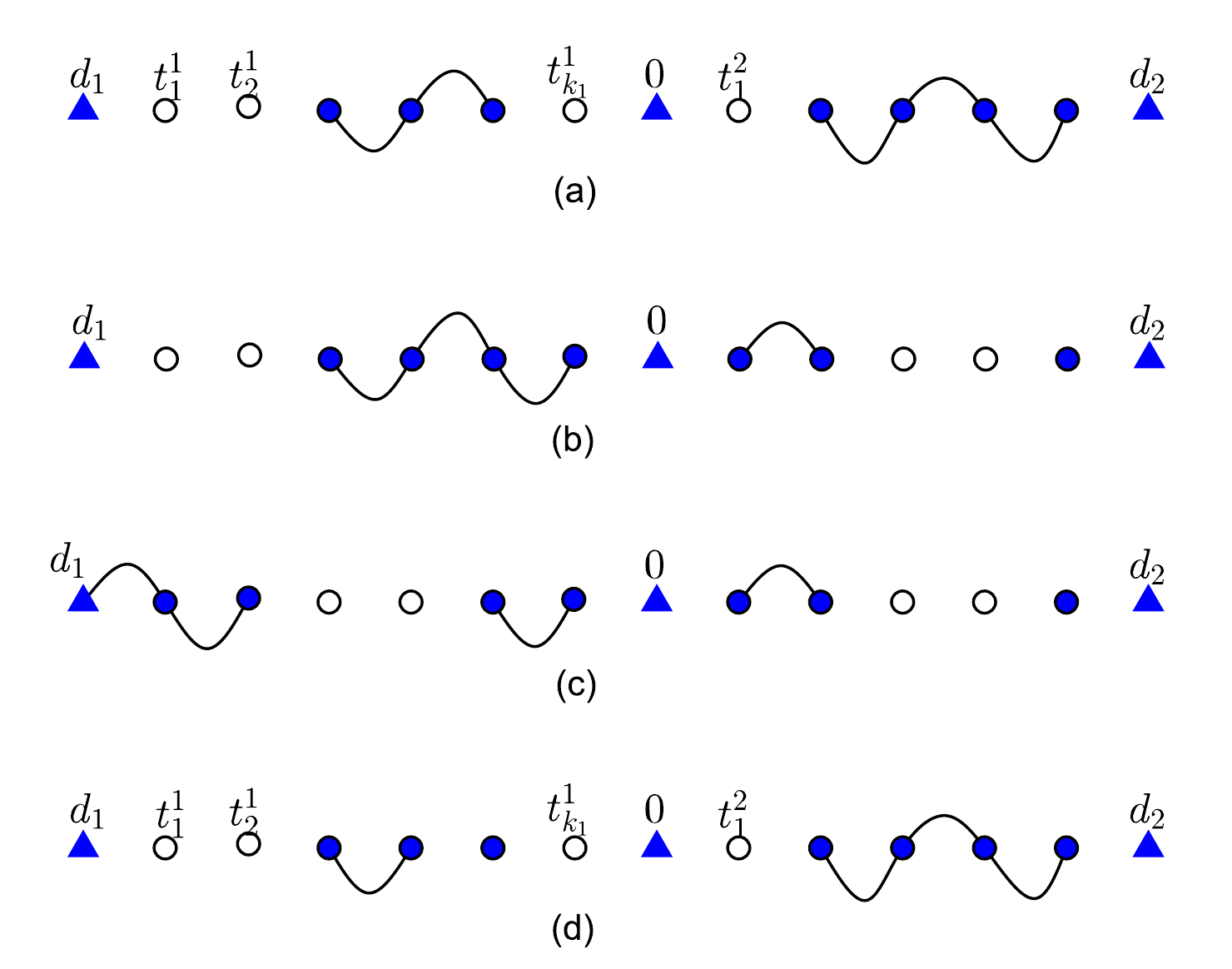}% \hspace{0.5cm}
\end{picture} 
}
%\end{picture}
%\put(8.8,0){\makebox(0,0)[cc]{$B$}}
\end{center}
\caption{ Illustrations of disassembling: (a) first step: $S_1$ and $S_2$ are separated by one sub-path only; (b) an example when $S_1$ and $S_2$ are separated by two sub-paths and a depot; (c) depot and the first subpath are considered as one customer; (d) modified first step: a subpath between $S_1$ and $S_2$ (case (a)) is partitioned into a sub-path and a single customer to keep the size of the new 2VRP fixed.
}
\label{fig:disassembling2}
\end{figure}

\paragraph{Details of implementation.} In the follow up computational experiments we used a modification of the approach described above. The parameters of the algorithm are the size of the subsets $s$ and step $l$. To keep the size of the small 2VRP fixed and defined by only these two parameters, we implement the disassembling procedure as described below.
We refer to Figure \ref{fig:disassembling2} in our explanations of various steps (and possibilities) for disassembling. Figure \ref{fig:disassembling2}(a) illustrates the first step of the disassembling process as was described above (with parameter $s=2$). Sets $S_1$ and $S_2$ are separated by one sub-path in this case. The size of the new small 2VRP  is $2s+5$.

In the illustrations we set step $l=2$.  Fig.~\ref{fig:disassembling2}(b)) illustrates the outcome of disassembling the tour on the second step. Notice that depot $0$ was already removed from the tour, therefore setting $l=2$ yields the position of $S_2$ as shown in the figure.  There are two sub-paths between sets $S_1$ and $S_2$, and the size of the new 2VRP problem is $2s+6$. 

Consider the step when $S_1$ and $S_2$ are chosen as shown in Fig.~\ref{fig:disassembling2}(c). If the first sub-path did not contain the depot, the size of the problem would have been $2s+7$.  On the implementation step, it was convenient to keep the size of the problem fixed at $2s+6$. Therefore it was decided to ``glue" the first sub-path with the depot and have it as the depot in the new problem as shown in the figure. 

If subsets  $S_1$ and $S_2$ are separated by a single path (for example, on the first step of disassembling), it was decided to consider the last node in the sub-path as a second sub-path: in this case the problem with $2s+5$ customers becomes the problem with $2s+6$ customers (compare Fig.~\ref{fig:disassembling2}(a) and Fig.~\ref{fig:disassembling2}(d)). 

We will use notation $H(s,l)$ for a heuristic from the family of heuristics described above, where the size of the subsets is $s$: $|S_1|=|S_2|=s$, and the step for moving the subsets is $l$.

%%%%%%%%%%%%%%%%%%%%%%%%%%%%%%%%%%%%%%%%%%%%%%%%%%%%%%%%%%%%%%%%%%%%%%%%%%

\subsection{Varieties of the VRP covered by the framework.} Below we illustrate the  advantages of the suggested dynamic programming approach by listing some types of the VRPs that can be tackled by the approach proposed.
\begin{itemize}
\item {\bf Arc routing.} Our model made no difference between the classic capacitated VRP, where the customers are represented by one node in a road network, and the arc routing VRP, where the customers are the streets/arcs in the network
(see \cite{Wohlk,CP}).

\item {\bf Heterogeneous fleet.} The dynamic programming recursions take into account individual characteristics of the vehicles, so both homogeneous and heterogeneous fleets (see \cite{BaldachiVigo} and recent survey \cite{KBJL})  can be managed.

\item {\bf Multi-depot and open VRPs.} Incorporation of the multi-depot feature  into the model is straightforward. The reader is referred to recent papers \cite{MTmultiD,KW} for the specifics of the multi-depot VRP. For the open VRP (see references in recent papers \cite{LJ,LSHD}), it is enough to introduce a dummy depot with zero distances to this depot from all customers.

\item {\bf Tight capacity constraint.} For some instances of the VRP, the main difficulties lie in packing all goods into a bounded number of vehicles. Since the dynamic programming approach enumerates all possible subsets, the ``bin packing"/loading part of the VRP is resolved at the same time as the routing part. See arguments in \cite{CGS} for the benefits of integrating loading and routing. It is easy to see that the framework can incorporate  more complicated packing constraints, e.g. two-dimensional loading constraints \cite{LZZ}, by solving the corresponding packing sub-problems on each step of calculations. 

\item {\bf Fixed items in a vehicle.} In some VRPs it is important to allocate customers to particular vehicle visits in advance. We refer to these customers as \emph{fixed items} in a vehicle. This feature can easily be added to the dynamic programming recursions. Fixed items can be useful, e.g. with multiple visits of customers (see Section \ref{sec:balanced} in this paper). Another example is the so-called site-dependent VRP \cite{BaldachiVigo} - some customers can be served only by a specific type of vehicles (so-called \emph{docking} constraints \cite{CGRC}). An interesting case study with fixed customers was described in \cite{DGHL08}. An Austrian red cross considered introducing two tiers for a blood delivery service. Urgent delivery (with a higher price) is delivery within one day, and standard delivery (at a lower price) is delivery on the second day. Hospital customers for the current day are known, while the next day's customers are unknown and only the probabilities of requests can be evaluated. So, on each day a dispatcher knows undelivered requests from yesterday, requests which arrived today, and probable requests for tomorrow. The requests from yesterday are urgent and have to be delivered today - hence a fixed allocation of these customers to today's route (it is assumed here that the delivery is done by one vehicle). A sample of (probable) requests for tomorrow should be fixed for tomorrow's delivery. Today's requests are flexible and can be allocated to either vehicle, however they will be charged different prices.

\item {\bf Penalties for wrong day deliveries.} In the previous paragraph we mentioned charges/costs for deliveries in different vehicles/days. Another example is given in \cite{CRW15}, where penalties for wrong day deliveries were introduced. These can easily be incorporated into our model by changing  left/right lengths of intervals (assuming that costs of travelling and penalties are measured in the same monetary units).

\item {\bf Cumulative VRPs.} In the \emph{cumulative} VRP the objective is to minimise the sum of arrival times to all customers; this problem is also known as the \emph{latency} problem (see references in recent papers \cite{CSW,LW,RAP}). It is easy to see that the dynamic programming approach can be adapted for this type of objective function.

\item {\bf Weight or time dependent travel costs.}  There are some practical situations when the  travel costs depend on the load of the vehicle (see  \cite{ZTK,ZQZL} and references there) or on the time when the vehicle travels (\cite{ECT,TGJL}). It can easily be seen that the dynamic programming recursions above allow this type of calculation to be incorporated into the recursions.

\end{itemize}

%%%%%%%%%%%%%%%%%%%%%%%%%%%%%%%%%%%%%%%%%%%%%%%%%%%%%%%%%%%%%%%%%%%%%%%%%
%%%%%%%%%%%%%%%%%%%%%%%%%%%%%%%%%%%%%%%%%%%%%%%%%%%%%%%%%%%%%%%%%%%%%%%%%
\begin{figure}
\unitlength=1cm
\begin{center}
%\begin{picture}(12,7)
%\put(-2,0.5){\begin{picture}(5,4)
%\includegraphics[scale=1]{Instance2.pdf}% \hspace{0.5cm}
%\end{picture} }
%\put(1.5,0.2){\makebox(0,0)[cc]{$(a)$}}
%\put(6,0.5)
{\begin{picture}(6.5,6)
\includegraphics[scale=1]{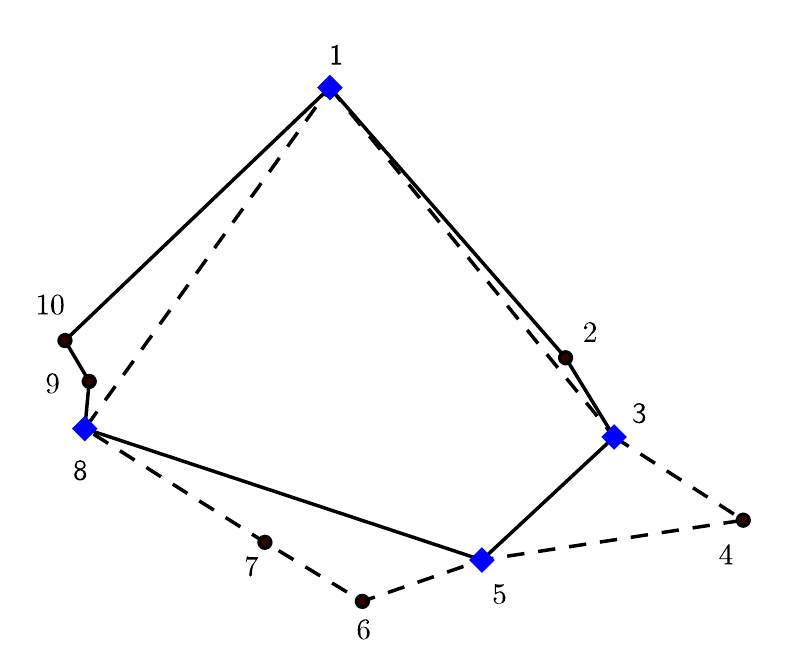}% \hspace{0.5cm}
\end{picture} }
%\put(10,0.2){\makebox(0,0)[cc]{$(b)$}}
%\end{picture}
\end{center}
\caption{Illustration of the definition of the balanced 2TSP; 
customers $1$, $3$, $5$, and $8$ are visited in two periods.}
\label{fig:2pTSP}
\end{figure}

\section{An application of the framework}
In this section we provide evidence of computational efficiency of the proposed framework. 
As a test problem we have chosen a 2-period balanced travelling salesman problem and a set of the benchmark test problems from Bassetto \& Mason \cite{BassettoM11}.
In their algorithms, Bassetto \& Mason used powerful exact techniques and even used a visualisation and human interventions for improving solutions. We believe that we have chosen a strong competitor to test the potential of the suggested framework.

\subsection{A 2-period travelling salesman problem}\label{sec:balanced}

Given $ m $ customers to be visited in each of the two periods and $ n $ customers to be visited once in either of the periods, and the distance matrix between the customers,  the 2-period travelling salesman problem (2TSP) asks for tours for the two periods with the minimal total distance travelled. 

Butler, Williams, and Yarrow \cite{ButW97} considered a practical application of the 2TSP in milk collection in Ireland. They applied what they called a ``man-machine method'', combining an integer programming technique with a human being intervention for identifying violated constraints. 

The 2TSP can be modelled as a special case of the 2VRP  as follows. For each of the customers to be visited twice, an identical copy of the customer is created. The identical customers are allocated then to the different vehicles. This allocation to the vehicles is fixed.  The 2VRP now has $n+2\times m$ customers: each vehicle has to serve $m$ fixed customers, $n$ customers to be served by only one of the two vehicles. 
%The so defined 2VRP (with fixed items) can be found as a sub-problem in more general period VRP \cite{FST}

In the \emph{balanced} 2TSP (2VRP) an additional constraint demands that the number of customers visited in each period (each vehicle) differs by no more than 1.
Bassetto and Mason \cite{BassettoM11} considered a balanced 2TSP in the Euclidean plane. In this variant of the problem the customers' locations are points in the Euclidean plane, and the costs of travelling between the customers are standard Euclidean distances. 
Figure~\ref{fig:2pTSP} illustrates a solution to the Euclidean balanced 2TSP with $10$ customers. Four out of ten customers in the example are visited in $2$ periods (the depot is considered as a customer here).

%Our framework will be tested on a version of the balanced 2TSP and the benchmark instances considered by Basseto \& Mason \cite{BassettoM11}. 

A short summary of the approaches used in \cite{BassettoM11} is as follows. First a TSP tour on the set of all customers, called a general tour (GT), is constructed. The GT is used to obtain a partition of customers into two subsets visited in two periods. The initial partition is improved by applying decision rules motivated by geometry (e.g. removal of crossing edges).
For each period an optimal TSP tour is constructed by applying an exact TSP algorithm (see \cite{ABCC}, Chapter 16,  and \cite{Concorde}). Three algorithms A1, A2, and A3 are suggested and tested on the benchmark instances.
Since the test instances are defined by the Euclidean coordinates, a visualisation of solutions is an option. The authors used the visualisation and human intervention to improve some of the solutions. Columns labelled as ``PC"  in the tables in the Appendix refer to the best solutions 
found by algorithms A1 and A3 from \cite{BassettoM11}. Columns ``PC+Manual" refer to the best  solutions obtained after manual improvements (the exceptions are the results for instances $I_5$, $I_6$, and $I_{10}$ the best solution for which were found by algorithm A2).

\subsection{Computational experiments}

The set of benchmark instances from \cite{BassettoM11} contains 60 randomly generated instances with 48 customers. The set of these instances is divided into three subsets with a different number of customers to be visited in the two periods ($8$, $12$, and $24$ customers).  We used the same set of instances in our computational experiments. However, we used only distance matrices, without using in our algorithms any knowledge on the sets of coordinates in the instances.

%The authors considered three heuristics to transform GT into a partition into 

We used the framework discussed in the previous sections in the following algorithmic setting:
\begin{tabbing}
1111\=1111\=1111\=\kill
{\bf Algorithm for the 2VRP \{ }\\
\>{\bf repeat}  $n (=48)$ times:\\
\>\>Generate a random allocation of $n$ customers into two TSP tours;\\
\>\>Insert into each tour the missing fixed customers to be visited by the two vehicles;\\
\>\>Apply a tour improvement TSP heuristic of Carlier \& Villon \cite{CV87} to improve the tours;\\
\>\>\bf{repeat}\\
\>\>\>Apply dynamic programming heuristic $H(s,l)$ for improving the partition into the  tours;\\
\>\>\>Apply a tour improvement TSP heuristic of Carlier \& Villon \cite{CV87} to improve the tours;\\
\>\>{\bf while} (improvement found);\\
{\bf \}}
\end{tabbing}

\begin{figure}
\unitlength=1cm
\begin{center}
{\begin{picture}(10,6.5)
%\includesvg{Histogram.svg}% \hspace{0.5cm}
%\includegraphics[scale=1]{Instance1}% \hspace{0.5cm}
%\includepdf[pages={1}]{hist_10bins.pdf}
\includegraphics[scale=0.5]{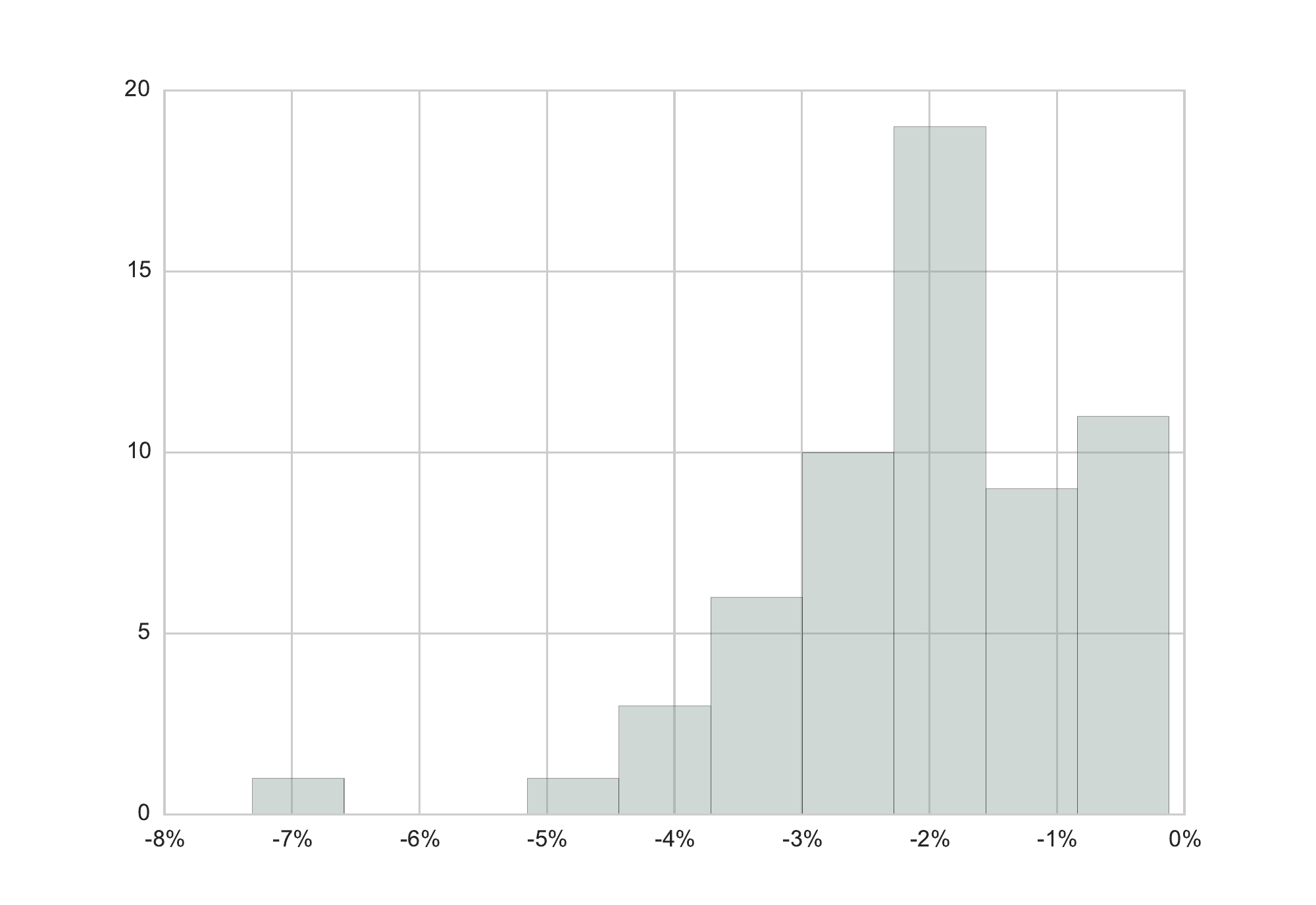}% \hspace{0.5cm}
%\includepdf{hist_10bins.pdf}
\end{picture} 
}
\end{center}
\caption{Histogram for the distribution of improvements achieved with $H(6,3)$ on all 60 instances.}
\label{fig:hist}
\end{figure}

In our computational experiments we used three heuristics: $H(3,1)$, $H(5,2)$, and $H(6,3)$. So the dynamic programming recursions were used for sub-problems with no more than 18 customers. We found that the results are much better if a multi-start strategy is used, therefore we repeated our search starting with randomly generated initial solutions. The number of repetitions in our experiments was chosen as $n=48$, however for some of the instances we tried $2n$ repetitions to improve the solutions (see comments in the appendix). Surprisingly random partitions into tours in our experiments yield better solutions than partitions obtained from a ``good" (general) TSP solution for all customers. We did not use an exact algorithm for finding TSP solutions. We used  Carlier \& Villon's \cite{CV87} TSP heuristic which is an $O(n^3)$ heuristic that finds  an optimal TSP tour in an exponential neighbourhood of tours. (The use of this particular heuristic is rather a matter of author's preferences. Notice that the neighbourhood searched by this heuristic contains at least $75\%$ of tours that can be obtained from a tour by applying the well-known $3opt$ heuristic, as proved in Theorem 5 in \cite{Neighborhoods}.
For the experiments we used a desktop computer with Intel i7-3770 3.40 GHz CPU, 16 GB of RAM, and GNU C++ compiler. The length of the solutions found are represented in the tables in the Appendix. In this section we provide a summary of the results obtained. 
 
Figure \ref{fig:hist} shows the improvement (measured in percent of achieved cost) that our technique yields using heuristic $H(6,3)$ for the dynamic programming search. Our results here are compared to the best solution found by algorithms A1 and A3 in \cite{BassettoM11}. It can be seen that all solutions have been improved, with the majority of improvements being  $2\%$ or better. By varying the size of the neighbourhoods searched in the dynamic programming heuristic, one can influence the computation time (and the accuracy). Tables \ref{table:summary1}-\ref{table:summary3} below summarise the outcomes of computational experiments for each of the settings used and for each type of instance. We compared our computational results with both the computer results (labelled as ``PC") and the results improved by manual intervention (labelled as ``PC+Manual") from \cite{BassettoM11}. 

\begin{table}
\begin{center}
\begin{tabular}{||l||c|c||c|c||c|c||}
\hline 
% \multirow{2}{*}
 &
  \multicolumn{2}{c||}{$H(3,1)$, $t_{mean}=27$ sec} & \multicolumn{2}{c||}{$H(5,2)$, $t_{mean}=172$ sec} & \multicolumn{2}{c||}{$H(6,3)$, $t_{mean}=415$ sec} \\ 
%\cline{2-7} 
\hline
Settings in \cite{BassettoM11}  & PC &PC+manual & PC & PC+manual & PC & PC+manual \\ 
%\hline 
Mean \% & -2.22\% & -0.51\% & -2.54\% & -0.83\% & -2.63\% & -0.93\% \\ 
 Best \%& -7.19\% & -1.75\% & -6.92\% & -2.65\% & -7.31\% & -2.77\% \\ 
Worst \%& +0.46\% & +0.86\% & +0.28\% & +0.93\% & -0.12\% & -0.00\% \\ 
Improved \#& 19/20 & 13/20 & 19/20 & 17/20 & 20/20 & 20/20 \\ 
\hline 
\end{tabular} 
\end{center}
\caption{Summary of results for instances with $8$ (out of $48$) nodes  visited in two periods.}
\label{table:summary1}
\end{table}

\begin{table}
\begin{center}
\begin{tabular}{||l||c|c||c|c||c|c||}
\hline 
% \multirow{2}{*}
 &
  \multicolumn{2}{c||}{$H(3,1)$, $t_{mean}=23$ sec} & \multicolumn{2}{c||}{$H(5,2)$, $t_{mean}=172$ sec} & \multicolumn{2}{c||}{$H(6,3)$, $t_{mean}=415$ sec} \\ 
%\cline{2-7} 
\hline
Settings in \cite{BassettoM11}  & PC &PC+manual & PC & PC+manual & PC & PC+manual \\ 
%\hline 
Mean \% & -1.76\% & -0.63\% & -2.14\% & -1.02\% & -2.17\% & -1.04\% \\ 
 Best \%& -3.81\% & -2.46\% & -3.93\% & -2.59\% & -3.93\% & -2.59\% \\ 
Worst \%& +0.30\% & +2.09\% & -0.53\% & +0.42\% & -0.53\% & +0.46\% \\ 
Improved \#& 19/20 & 17/20 & 20/20 & 18/20 & 20/20 & 19/20 \\ 
\hline 
\end{tabular} 
\end{center}
\caption{Summary of results for instances with $16$ (out of $48$) nodes  visited in two periods.}
\label{table:summary2}
\end{table}

\begin{table}
\begin{center}
\begin{tabular}{||l||c|c||c|c||c|c||}
\hline 
% \multirow{2}{*}
 &
  \multicolumn{2}{c||}{$H(3,1)$, $t_{mean}=17$ sec} & \multicolumn{2}{c||}{$H(5,2)$, $t_{mean}=131$ sec} & \multicolumn{2}{c||}{$H(6,3)$, $t_{mean}=299$ sec} \\ 
%\cline{2-7} 
\hline
Settings in \cite{BassettoM11}  & PC &PC+manual & PC & PC+manual & PC & PC+manual \\ 
%\hline 
Mean \% & -1.20\% & -0.58\% & -1.25\% & -0.64\% & -1.36\% & -0.75\% \\ 
 Best \%& -4.44\% & -1.50\% & -3.85\% & -1.52\% & -4.44\% & -2.17\% \\ 
Worst \%& -0.00\% & +0.24\% & -0.17\% & +0.24\% & -0.17\% & +0.16\% \\ 
Improved \#& 20/20 & 19/20 & 20/20 & 19/20 & 20/20 & 18/20 \\ 
\hline 
\end{tabular} 
\end{center}
\caption{Summary of results for instances with $24$ (out of $48$) nodes  visited in two periods.}
\label{table:summary3}
\end{table}

\section{Summary} In this paper we have presented a framework for small but rich VRPs. The framework can be used for a variety of VRP settings. The computational scheme which implements the framework permits an easy scaling,  hence varying computational time and the accuracy of the solutions found. Computational experiments for the balanced 2-period TSP have shown an impressive performance from the framework. Our next step would be performing extensive computational experiments to test the potential of the framework for other VRP settings.

\section*{Acknowledgements}
\nopagebreak
Vladimir Deineko acknowledges support
by the Center for Discrete Mathematics and Its Applications, University of Warwick. The author also thanks Tatiana Bassetto and Francesco Mason for providing the benchmark problems, Douglas Miranda and Alex Taylor for useful comments on an early version of the paper.

\appendix
\section {Appendix } \nopagebreak[4]
\begin{center}
\begin{tabular}{||c|c|c||c|c||c|c||c|c||}
%\multicolumn{9}{l}{\textbf{Results of computational experiments for instances $8$ out of $48$ nodes  48 nodes \cite{FrancisST09};}}\\
%\multicolumn{9}{l}{\textbf{8 nodes are visited in two periods.}}\\
\multicolumn{9}{l}{\textbf{Results for instances with $8$ out of $48$ nodes visited in two periods}}\\
\hline 
  \multirow{2}{*}& \multicolumn{2}{c||}{Solutions from \cite{BassettoM11}}& \multicolumn{2}{c||}{$H(3,1)$} & \multicolumn{2}{c||}{$H(5,2)$} & \multicolumn{2}{c||}{$H(6,3)$ } \\ 
\cline{2-3} \cline{4-5} \cline{6-9} 
Instance & PC & PC+manual &time (sec) & length&time (sec) & length &time (sec)& length \\ 
%Instance & Algorithms A1,A3& PC+manual &time & length&time & length &time & length \\ 
\hline 
$I_{21}$	& 25217	& 24937	& 26	& 24517	& 185	& \textbf{24493}	& 395	& 24517\\	
$I_{22}$	& 26996	& 26549	& 19	& 26413	& 141	& \textbf{26347}	& 352	& \textbf{26347}\\	
$I_{23}$	& 26476	& 26192	& 23	& 26222	& 168	& 26131	& 412	& \textbf{26050}\\	
$I_{24}$	& 26802	& 26038	& 30	& 26091	& 169	& 26057	& 369	& \textbf{26002}\\	
$I_{25}$	& 27728	& 27408	& 23	& 27053	& 150	& \textbf{26914}	& 391	& \textbf{26914}\\	
$I_{26}$	& 24348	& 24268	& 29	& 23894	& 186	& \textbf{23837}	& 463	& \textbf{23837}\\	
$I_{27}$	& 27335	& 26857	& 26	& 26435	& 181	& \textbf{26417}	& 422	& \textbf{26417}\\	
$I_{28}$	& 24679	& 24232	& 26	& 24006	& 162	& \textbf{23985}	& 408	& 24143\\	
$I_{29}$	& 26890	& \textbf{26466}	& 28	& \textbf{26466}	& 180	& \textbf{26466}	& 405	& \textbf{26466}\\	
$I_{30}$	& 24978	& 24200	& 32	& 24407	& 212	& 23915	& 549	& \textbf{23915}\\	
$I_{31}$	& 26266	& 26130	& 23	& 26130	& 147	& \textbf{26130}	& 369	& \textbf{26130}\\	
$I_{32}$	& 26360	& 26054	& 32	& 26067	& 196	& 26296	& 429	& \textbf{26032}\\	
$I_{33}$	& 26418	& 26418	& 28	& 26540	& 174	& 26493	& 419	& \textbf{26387}\\	
$I_{34}$	& 28733	& 27074	& 28	& 26666	& 166	& 26744	& 380	& \textbf{26633}\\	
$I_{35}$	& 25043	& 24587	& 27	& 24587	& 172	& 24526	& 434	& \textbf{24517}\\	
$I_{36}$	& 27103	& 26790	& 26	& 26321	& 179	& 26079	& 409	& \textbf{26079}\\	
$I_{37}$	& 25662	& 25123	& 24	& \textbf{25072}	& 153	& \textbf{25072}	& 432	& 25101\\	
$I_{38}$	& 26459	& 25709	& 26	& 25816	& 161	& \textbf{25588}	& 372	& \textbf{25588}\\	
$I_{39}$	& 27209	& 26994	& 27	& 26553	& 186	& 26314	& 474	& \textbf{26247}\\	
$I_{40}$	& 25416	& 24964	& 28	& 25081	& 170	& 24850	& 419	& \textbf{24824}\\	
\hline 
\end{tabular} 
\end{center}
\begin{center}
\begin{tabular}{||c|c|c||c|c||c|c||c|c||}
\multicolumn{9}{l}{\textbf{Results for instances with $16$ out of $48$ nodes visited in two periods}}\\
%\multicolumn{9}{l}{\textbf{16 nodes are visited in two periods.}}\\
\hline 
  \multirow{2}{*}& \multicolumn{2}{c||}{Solutions from \cite{BassettoM11}}& \multicolumn{2}{c||}{$H(3,1)$} & \multicolumn{2}{c||}{$H(5,2)$} & \multicolumn{2}{c||}{$H(6,3)$ } \\ 
\cline{2-3} \cline{4-5} \cline{6-9} 
Instance & PC & PC+manual &time (sec) & length&time (sec) & length &time (sec)& length \\ 
%Instance & Algorithms A1,A3& PC+manual &time & length&time & length &time & length \\ 
\hline 
$I_{1}$	& 33804	& \textbf{32556}	& 25	& 32566	& 154	& 32566	& 392	& \textbf{32556}\\	
$I_{2}$	& 30929	& 30929	& 23	& \textbf{30716}	& 150	& \textbf{30716}	& 379	& \textbf{30716}\\	
$I_{3}$	& 30596	& 30382	& 24	& 30280	& 157	& 29967	& 385	& \textbf{29954}\\	
$I_{4}$	& 28563	& 28441	& 23	& 28274	& 139	& 28260	& 337	& \textbf{28223}\\	
$I_{5}$	& 27323	& 27206	& 29	& \textbf{27177}	& 184	& \textbf{27177}	& 452	& \textbf{27177}\\	
$I_{6}$	& 33065	& 32396	& 22	& 31971	& 152	& \textbf{31890}	& 346	& \textbf{31890}\\	
$I_{7}$	& 32854	& 31861	& 24	& 31615	& 153	& \textbf{31606}	& 370	& 31637\\	
$I_{8}$	& 30850	& 30571	& 22	& \textbf{30215}	& 120	& 30266	& 328	& 30266\\	
$I_{9}$	& 34709	& 34024	& 19	& 33951	& 124	& \textbf{33911}	& 331	& \textbf{33911}\\	
$I_{10}$	& 31451	& 30867	& 25	& 30844	& 154	& \textbf{30660}	& 368	& \textbf{30660}\\	
$I_{61}$	& 27158	& 26934	& 22	& 26799	& 146	& \textbf{26644}	& 348	& \textbf{26644}\\	
$I_{62}$	& 27774	& 27619	& 24	& 27325	& 168	& \textbf{27248}	& 360	& \textbf{27248}\\	
$I_{63}$	& 25308	& 24960	& 21	& 24345	& 142	& \textbf{24313}	& 324	& \textbf{24313}\\	
$I_{64}$	& 27875	& \textbf{27285}*	& 19	& 27856	& 138	& 27399	& 377	& 27410\\	
$I_{65}$	& 27060	& 26888	& 20	& 27141	& 140	& 26872	& 373	& \textbf{26806}\\	
$I_{66}$	& 27677	& 27624	& 26	& 27135	& 169	& 27188	& 429	& \textbf{27131}\\	
$I_{67}$	& 30268	& 30203	& 21	& 29928	& 156	& \textbf{29841}	& 340	& \textbf{29841}\\	
$I_{68}$	& 28033	& 27923	& 24	& 27478	& 163	& \textbf{27425}	& 393	& 27447\\	
$I_{69}$	& 27958	& 27638	& 22	& 27621	& 154	& \textbf{27143}	& 383	& \textbf{27143}\\	
$I_{70}$	& 28483	& 28427	& 22	& 27865	& 157	& 27799	& 350	& \textbf{27790}\\	
\hline 
\end{tabular} 
\end{center}
Solution for instance $I_{64}$ was improved to $27236$ by $H(5,2)$ with $2\times 48$ random starts  (in $280$ seconds).

\begin{center}
\begin{tabular}{||c|c|c||c|c||c|c||c|c||}
\multicolumn{9}{l}{\textbf{Results for instances with $24$ out of $48$ nodes visited in two periods}}\\
%\multicolumn{9}{l}{\textbf{24 nodes are visited in two periods.}}\\
\hline 
  \multirow{2}{*}& \multicolumn{2}{c||}{Solutions from \cite{BassettoM11}}& \multicolumn{2}{c||}{$H(3,1)$} & \multicolumn{2}{c||}{$H(5,2)$} & \multicolumn{2}{c||}{$H(6,3)$ } \\ 
\cline{2-3} \cline{4-5} \cline{6-9} 
%Instance & Algorithms A1,A3& PC+manual &time & length&time & length &time & length \\ 
Instance & PC & PC+manual &time (sec) & length&time (sec) & length &time (sec)& length \\ 
\hline 
$I_{41}$	& 30253	& 30147	& 16	& 30100	& 117	& \textbf{30064}	& 258	& \textbf{30064}\\	
$I_{42}$	& 33008	& 32020	& 17	& \textbf{31544}	& 124	& 31738	& 284	& \textbf{31544}\\	
$I_{43}$	& 31500	& 31500	& 18	& 31200	& 140	& \textbf{31166}	& 311	& 31194\\	
$I_{44}$	& 30313	& 30170	& 21	& 29812	& 156	& \textbf{29757}	& 335	& \textbf{29757}\\	
$I_{45}$	& 27986	& 27857	& 17	& \textbf{27780}	& 128	& \textbf{27780}	& 310	& \textbf{27780}\\	
$I_{46}$	& 30073	& 30013	& 20	& \textbf{30013}	& 148	& \textbf{30013}	& 328	& \textbf{30013}\\	
$I_{47}$	& 32106	& 32106	& 17	& \textbf{31704}	& 133	& 31735	& 299	& 31728\\	
$I_{48}$	& 31004	& 30942	& 19	& 30478	& 149	& \textbf{30471}	& 337	& \textbf{30471}\\	
$I_{49}$	& 33663	& 33185	& 17	& \textbf{33173}	& 130	& \textbf{33173}	& 312	& 33197\\	
$I_{50}$	& 31266	& 31266	& 16	& 31266	& 117	& \textbf{31213}	& 259	& \textbf{31213}\\	
$I_{51}$	& 33722	& 33627	& 19	& 33358	& 146	& \textbf{33344}	& 322	& 33358\\	
$I_{52}$	& 32353	& 32280	& 15	& 32251	& 113	& 32251	& 274	& \textbf{32200}\\	
$I_{53}$	& 33287	& 33200	& 17	& 32943	& 118	& 32871	& 268	& \textbf{32726}\\	
$I_{54}$	& 31973	& 31600	& 18	& \textbf{31368}	& 149	& 31373	& 345	& \textbf{31368}\\	
$I_{55}$	& 33837	& \textbf{33507}*	& 16	& 33587	& 115	& 33587	& 292	& 33560\\	
$I_{56}$	& 29696	& 29476	& 16	& 29176	& 115	& 29156	& 266	& \textbf{28835}\\	
$I_{57}$	& 31954	& 31640	& 18	& \textbf{31427}	& 133	& \textbf{31427}	& 299	& \textbf{31427}\\	
$I_{58}$	& 30705	& 30246	& 18	& \textbf{30165}	& 140	& 30178	& 345	& \textbf{30165}\\	
$I_{59}$	& 31549	& 31549	& 19	& 31471	& 137	& \textbf{31223}	& 296	& \textbf{31223}\\	
$I_{60}$	& 32384	& 32317	& 15	& 32193	& 115	& \textbf{32137}	& 244	& 32140\\		
\hline 
\end{tabular} 
\end{center}
The best known value for instance $I_{55}$, which is $33507$, can be achieved within 620 seconds by our algorithm as well, if we set parameters as $s=6$ and $l=2$.

%%%%%%%%%%%%%%%%%%%%%%%%%%%%%%%%%%%%%%%%%%%%%%%%%%%%%%%%%%%%%%%%%%%%%%%%%
%%%%%%%%%%%%%%%%%%%%%%%%%%%%%%%%%%%%%%%%%%%%%%%%%%%%%%%%%%%%%%%%%%%%%%%%%

\end{document}